# COMPARISON OF NUMERICAL ALGORITHMS BASED ON ELEMENTARY AND MULLER'S BOUNDARY INTEGRAL EQUATIONS IN THE SCATTERING BY DIELECTRIC CYLINDERS

**Artem V. Boriskin [1], Svetlana V. Boriskina [2]**

[1] Institute of Radio-Physics and Electronics NASU, ul. Proskury 12, Kharkov 61085, Ukraine
[2] G. Green Institute for Electromagnetic Research, University of Nottingham, NG7 2RD, UK
E-mail: a_boriskin@yahoo.com

*Abstract* - We consider boundary integral equations (BIEs) met in the scattering by dielectric cylinders and compare numerical algorithms based on elementary and Muller's BIEs near the "numerical resonances". A procedure of partial removal of well-known defect of elementary BIEs related to the loss of their unique solvability is discussed.

Elementary BIEs are obtained if the fields are presented in terms of single or double-layer potentials over the contour of a homogeneous dielectric scatterer. If a combination of the potentials with a proper choice of the coefficients is used then the equations take a specific form known as Muller's BIE [1]. The main advantage of the latter is that they are free from the defect of the loss of unique solvability that is intrinsic for elementary BIE. This loss takes place at a countable set of real positive values of the wavenumber which are the eigenvalues of the interior Dirichlet or Neumann boundary-value problem [2]. Besides, the same defect values are the poles of the condition number of the exact infinite matrix equation obtained after the discretization of the BIE. Truncation of the matrix equation as well as finite accuracy of the matrix elements calculation transforms the defect points to finite intervals with unpredictable computational error, so-called "numerical resonances" (NR). The width of NR depends on the discretization scheme and the matrix filling accuracy. This defect makes elementary BIE almost inapplicable for the frequency analysis. Nevertheless, the algorithm properties can be improved if one considers the following two factors: a) the condition number suffers a significant jump near a defect value of the frequency parameter, b) the spectrum of defect frequencies is determined by the scatterer geometry but not the permittivity. Keeping this in mind, the following procedure for an accurate interpolation to the "defective" regions can be proposed: a) define such regions by some level of the condition number, b) make calculation in the neighboring (shifted) region for slightly changed $\varepsilon$, c) replace the sought function in the "defective" region with its values obtained in the shifted region. Fig. 1 shows forward and backward scattering cross-sections for a smooth homogeneous dielectric cylinder illuminated by a plane E-wave for two different but close values of $\varepsilon$. NR at the fixed value of the normalized frequency parameter $ka$ is well seen as well as a shift of the physical resonances of the scatterer. Fig. 2 shows the same characteristic as Fig. 1 near another NR calculated with the algorithms based on elementary and Muller's BIE. As expected, the defect does not appear for the Muller BIEs.

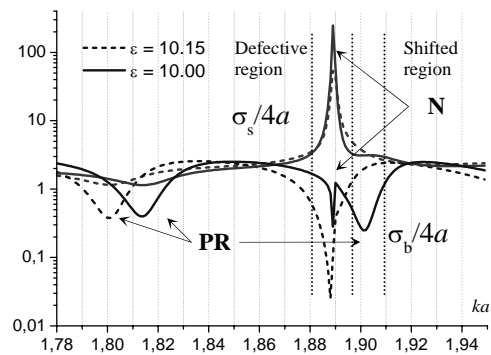

Fig. 1. Identification of numerical resonances (NR) and physical resonances (PR).

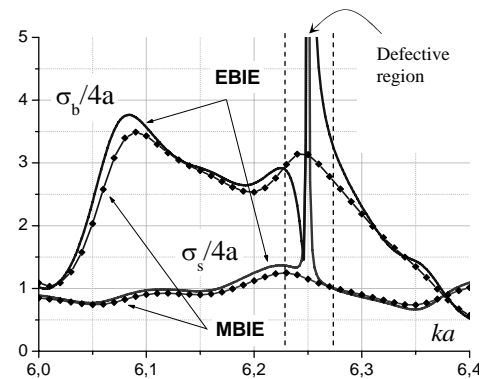

Fig. 2. Comparison between elementary and Muller's BIE.

Nevertheless, we admit that the best way for overcoming the defect of deceivingly simpler elementary BIEs is to switch to the Muller BIEs.